\documentclass{article}
\usepackage{graphicx}
\usepackage{subfig}
\usepackage{multirow}
\usepackage{amsfonts}
\usepackage{xcolor}

\newtheorem{theorem}{Theorem}

\usepackage{fancyhdr}
\begin{document}

\title {The observed Fisher information attached to the EM algorithm, illustrated on Shepp \& Vardi's estimation procedure for positron emission tomography}

\author {
Isaac Meilijson \\
{\em School of Mathematical Sciences} \\
{\em Tel Aviv University, Tel Aviv 6997801, Israel} \\
{\em E-mail: \tt{isaco@tauex.tau.ac.il}} \\
}


\maketitle

\begin{abstract}

\noindent The Shepp \& Vardi (1982) implementation of the EM algorithm provides a point estimate of the PET scan tumor radiotracer distribution. The current study
presents a closed-form formula of the observed Fisher information for Shepp \& Vardi PET scan estimator. A simpler approximate formula is presented as well, evaluated from the empirical projection data, before the application of the EM algorithm.

\noindent Keywords: PET scan, EM algorithm, Fisher information matrix, standard errors.
\end{abstract}

\pagenumbering{arabic}

\section{Introduction} \label{introduc}

The tissue for PET (positron emission tomography) purposes is divided into voxels $b$. The tumor radiotracer distribution $\lambda$ is a non-negative function on the tissue. Concomitant with radiotracer decay, positron/electron annihilation events produce at the voxels two high energy photons, travelling in opposite directions, which can be detected coincidently by pairs of detectors. Let the projection data $n_d^*$ stand for the total number of such photon pairs detected by the pair $d$ of detectors. These projection data $n_d^*$, also termed sinogram, are the statistical observations. Let $p(b,d)$ be the system matrix, the probability that an annihilation event at voxel $b$ is detected by the detector pair $d$. These probabilities $p$ are pre-processed data mostly hard-wired into the program, considered known when analyzing the patient's $n^*$-data.

Shepp \& Vardi (1982) took $\lambda_b$ to be the expected number of annihilation events in voxel $b$ during the entire acquisition period. Since the current study deals with the standard errors of the radiotracer estimates, $\lambda$ will be given a unit-time interpretation. During a unit time (arbitrarily set) there is a Poisson distributed number of annihilation events in voxel $b$, with mean $\lambda_b$. These annihilation events are (lost or) detected by the detector pairs to collect the projection data $n^*$, aggregated over the acquisition period of $T$ unit times. This will better capture and enhance the dependence of standard errors on acquisition time.

In principle, the regression equation $p' \lambda \sim {{n^*} \over T}$ could have provided a regression estimate of $\lambda$, but matrix size prevents the application of this obvious first idea. The EM-algorithm (expectation-maximization) is a practical, implementable way to arrive at a version of the solution. Proceeding with the weighted regression idea further, the covariance matrix of the right hand side is diagonal, and due to the Poisson nature of $n^*$, the $d$-entry can be approximated by ${{n_d^*} \over {T^2}}$. Hence, letting $D$ be the diagonal matrix with ${T \over n_d^*}$ along the diagonal and $1$ a vector of $1$'s, $\lambda$ would have been estimated by Least Squares as $\hat{\lambda}_{LS}=(p D p')^{-1} 1$. The covariance matrix of this estimate $\hat{\lambda}_{LS}$ of $\lambda$ is approximately $\hat{C}={1 \over T}(p D p')^{-1}$.

Unlike the evaluation of $\hat{\lambda}_{LS}$, that seems to require full inversion of $p D p'$, the evaluation of small minors of its covariance matrix may be accomplished locally, as long as dependence of the entries of $\hat{\lambda}_{LS}$ decay fast enough with voxel distance.

The purpose of the current study is to introduce an exact formula for the Fisher information matrix, to be evaluated via the EM algorithm. Since the MLE (maximum likelihood estimator) $\hat{\lambda}_{MLE}$ and the Least Squares estimator $\hat{\lambda}_{LS}$ are bound to be very similar, the observed Fisher information matrix is bound to be close to $T p D p'$. These similarities, apparent in Theorem 1 listed below, will be illustrated. In practice, it is proposed to keep estimating $\lambda$ by means of the EM algorithm, and estimate the covariance matrix in advance, via $p D p'$, built upon the data $p$ and $n^*$, without incorporating algorithmic changes into the Shepp \& Vardi program.

Covariance estimation will lead to the possibility of providing confidence intervals, design the exposure time $T$, calculate a window size for a moving average of the radiotracer distribution with small enough standard errors, and test hypotheses on relative severity of sections of the tumor.

What follows is a reminder of the Shepp \& Vardi method. Let $\lambda_b$ be the true tumor radiotracer value at voxel $b$ and let $\lambda_b^{(\infty)}=\hat{\lambda}_b$ be its MLE, maximum likelihood estimate, (or close enough to it, $\lambda_b^{(t)}$, at iteration $t$), obtained from the Shepp \& Vardi implementation on PET scan of the EM-algorithm (Expectation-Maximization) method of Dempster, Laird and Rubin (1977). For each detector pair $d$, $g^{(t)}_d=\sum_b \lambda_b^{(t)} p(b,d)=(p' \lambda^{(t)})_d$ is the expected sinogram estimate (per unit time) at iteration $t$.

Letting $q_b=1-\sum_d p(b,d)$ be the non-detection probability at voxel $b$, the general EM-algorithm iterative step is

\begin{equation} \label{EMstep}
\lambda_b^{(t+1)}=\lambda_b^{(t)}[q_b + {1 \over T} \sum_d p(b,d) {{n_d^*} \over {g^{(t)}_d}}]
\end{equation}
where $T=1$ and $q \equiv 0$ in the original Shepp \& Vardi formulation, where the authors argue (from splitting properties of Poisson processes) that if the sub-stochastic matrix $p$ is replaced by the stochastic matrix $p^*$ with $p^*(b,d)={{p(b,d)} \over {1-q_b}}$ and corresponding MLE estimate $\hat{\lambda}^*$, then $\hat{\lambda}$ with $\hat{\lambda}_b={{\hat{\lambda}_b^*} \over {1-q_b}}$ is the same as the MLE arrived at via (\ref{EMstep}). The two methods are obviously equivalent under the Least Squares version. While the two arrive at the same answer under MLE too, the Shepp \& Vardi stochastic-matrix modification is not only "simpler" as claimed, it is actually faster: reading (\ref{EMstep}) as $\lambda_b^{(t+1)}=\lambda_b^{(t)}[q_b + (1-q_b){1 \over T} \sum_d p^*(b,d) {{n_d^*} \over {g^{(t)}_d}}]$, the $p$-method advances in one iteration a proportion $1-q_b$ of what the $p^*$-method does.


The MLE is a point estimate of the true radiotracer distribution $\lambda$, as derived by Dempster, Laird and Rubin (1977) generally and applied to PET scan by Shepp and Vardi (1982), without error assessment, standard errors and confidence intervals. There is literature on bootstrap-based standard errors for PET scan (Markiewicz et al, 2014), and methods based on the assumption that $\lambda$ is smooth enough (Markiewicz et al, 2012).

On the other hand, there is literature on evaluating the observed Fisher information matrix based fully or partially on EM terms (Louis 1982, Meilijson 1989, Jamshidian and Jennrich 2000, Xu, Baines and Wang 2014), but it has not focused so far on PET scan.

The Fisher information matrix is (minus) the expectation of the Hessian of the logarithm of the likelihood function. The observed Fisher information is (minus) the Hessian itself, evaluated at the MLE of the parameter. It is the most commonly applied estimate of the Fisher information.

The current report illustrates on PET scan the methods above termed {\em EM-aided differentiation} in Meilijson (1989) and {\em numerical differentiation of the score} NDS in Jamshidian and Jennrich (2000) and Xu, Baines and Wang (2014), to identify the observed Fisher information matrix of the incomplete data wherever the EM algorithm is used to find MLE. As will be seen, this generally numerical method leads for PET scan to a closed-form formula. Since the estimated Fisher information is the actual observed Fisher information had the incomplete data been modelled, it is non-negative definite.

Under regularity conditions met by the Shepp and Vardi Poisson model, the inverse of the Fisher information matrix is the (asymptotic) covariance matrix of the MLE $\hat{\lambda}$. The square roots of the diagonal are estimates of its standard errors, and its off-diagonal terms estimate correlation coefficients between estimated radiotracer values at different voxels. As stated above, this allows error analysis and hypothesis testing of aggregations of voxels as well.

\section{EM-aided differentiation} \label{SecEMaided}

Here is a concise review of incomplete data terms. Using Meilijson (1989) as reference, the incomplete data log likelihood $\log(f_Y(y ; \theta))$ is represented as $\log(f_Y(y ; \theta))=Q(\theta,\theta_0)-V(\theta,\theta_0)$, where $Q$ is the working tool and $V$ has the property that for any fixed $\theta_0$, its maximal value over $\theta$ is attained at $\theta=\theta_0$. Hence, if $Q(\theta,\theta_0)>Q(\theta_0,\theta_0)$ then $\log(f_Y(y ; \theta)) > \log(f_Y(y ; \theta_0))$. This means that (under the proper regularity conditions) the EM iterative process ${{\partial Q(\theta,\theta_t)} \over {\partial \theta}}|_{\theta=\theta_{t+1}}=0$ increases the value of $\log(f_Y(y ; \theta))$ and if it converges, the limit is a stationary point of $\log(f_Y(y ; \theta))$. But, by the property of $V$ stated above, differentiation of $Q$ at the current iterate $\theta_0$ is meaningful too (Fisher, 1925).

\begin{equation} \label{Fisher25}
{{\partial Q(\theta,\theta_0)} \over {\partial \theta}}|_{\theta=\theta_0} = {{\partial \log(f_Y(y ; \theta))} \over {\partial \theta}}|_{\theta=\theta_0}
\end{equation}

Although incomplete data work circumvents the need to model the incomplete data density $f_Y(y ; \theta)$, its (vector) score function $S(\theta_0)={{\partial \log(f_Y(y ; \theta))} \over {\partial \theta}}|_{\theta=\theta_0}$ is a by-product of EM methodology. The observed Fisher information is (minus) the Jacobian, matrix of derivatives of the score function, Hessian of $\log(f_Y(y ; \theta))$ evaluated at the MLE $\hat{\theta}$ of $\theta$. Meilijson (1989), Subsection 3.4, EM-aided differentiation, and subsequently Jamshidian and Jennrich (2000) and Xu, Baines and Wang (2014), propose to evaluate this matrix numerically, by perturbing $\hat{\theta}$ by $\pm \epsilon$, for a small $\epsilon>0$, one coordinate at a time (using Kronecker $\delta_i$, a vector of all zeros except a single $1$ at coordinate $i$) to evaluate the two corresponding vector values of the score function. Their scaled differences ${{S(\theta+\epsilon \delta_i)-S(\theta-\epsilon \delta_i)} \over {2 \epsilon}}$
are numerical approximations of the sought after second derivatives, comprising the $i$'th row of the observed Fisher information corresponding to the $\epsilon$-perturbed $i$'th coordinate. However, for PET scan the current study identifies in closed-form the limit of these numerical second derivatives as $\epsilon \downarrow 0$.

\section{The Fisher information of PET scan} \label{FisherPET}

The observed Fisher information matrix has been defined above as (minus) the Hessian of the log likelihood function. The incomplete data log likelihood function is not evaluated by incomplete data methods. However, based on EM-aided differentiation (Section \ref{SecEMaided}), Theorem \ref{Fisherinfo} provides its Hessian (matrix of second derivatives) at arbitrarily chosen points. In principle it could speed up convergence by replacing the EM step by a Newton-Raphson step, but the computational cost may make this switch impractical. It is proposed to apply it at the point estimate arrived at by the EM algorithm, to estimate the Fisher information matrix. Experience will show the extent to which the simpler, pre-algorithmic covariance estimate $\hat{C}$ intended for design, is accurate enough for final analysis.

A diagonal matrix $D$ was defined in the Introduction in the context of the Regression solution. A modified version of this matrix plays a similar role under MLE. Let $D^{(t)}_d={1 \over T} {{n^*_d} \over {(g^{(t)}_d)^2}}$ and let ${\cal D}^{(t)}$ be the diagonal matrix with $D^{(t)}$ along the diagonal.
As can be inferred from (\ref{EMstep}), the radiotracer distribution MLE $\lambda^{(\infty)}$ (implicit in $g^{(\infty)}$) satisfies the identity $\sum_d  p(b,d) {{n_d^*} \over {g_d^{(\infty)}}}=T(1-q_b)$ for each voxel $b$. The Fisher information is built on similar terms, as the next Theorem states.

\begin{theorem} \label{Fisherinfo}

\{i\} The general EM-algorithm iterative step is given by (\ref{EMstep}).

\{ii\} The $(b_1,b_2)$ entry of (minus) the Hessian of the incomplete data log likelihood function is $I^{(t)}(b_1,b_2) = T \sum_d D^{(t)}_d p(b_1,d) p(b_2,d)$. I.e., $I^{(t)}=T p {\cal D}^{(t)} p'$. For large $t$, this is the observed Fisher information.

\end{theorem}

\noindent {\bf Remark}. The Fisher information matrix is non-negative. Hence, the radiotracer estimates have a tendency towards negative correlations, yielding less noisy moving averages.

\noindent {\bf Accuracy of Fisher information estimation}.
If the incomplete data log likelihood function is quadratic enough near the MLE $\lambda^{(\infty)}$ of the radiotracer distribution $\lambda$, the Hessian is nearly constant in its vicinity, and estimation of the Fisher information should then be more stable than the estimation of $\lambda$ itself. This may allow to use the same formula for estimation of the Fisher information matrix under regularization modifications of the EM algorithm. The simpler estimate of the Fisher information announced in the Introduction, based directly on the data $p$ and $n^*$, will be presented again and compared in the sequel.

\bigskip

\noindent {\bf The inverse of Fisher information}.
The covariance matrix of the MLE is estimated by the inverse of the Fisher information matrix, the Cr\'{a}mer-Rao bound. Large matrices can't be inverted in practice. The inverse of a minor of the Fisher information matrix provides the covariance matrix of the MLE of the corresponding parameters, as if all other parameters are {\em known} to be equal to their MLE estimates. This is a common method to approximate standard errors in spatial data and Kriging interpolation, typically from below. This issue will need detailed study in application.



\bigskip

\noindent {\bf Proof of Theorem 1}. The Shepp \& Vardi (1982) PET scan complete data likelihood is

\begin{equation} \label{complik}
f(x,t ; \lambda)=\prod_b {{e^{- T \lambda_b} (T \lambda_b)^{x_b}} \over {x_b!}} \prod_b \prod_d p(b,d)^{t_{b,d}}
\end{equation}
where $t_{b,d}$ is the sinogram count in the detector pair $d$ originating in annihilation events in voxel $b$, and $x_b=\sum_d t_{b,d}$. The sinogram $n_d^*=\sum_b t_{b,d}$ is viewed now as augmented by an additional detector $d_0$ that keeps track of undetected annihilation events. The sub-stochastic system matrix $p$ becomes stochastic with the addition of $d_0$, with $p(b,d_0)=q_b$. As the projection data $n_{d_0}^*$ is unobserved, it is part of the complete data but not of the incomplete data.

The incomplete data likelihood, a massive addition of (\ref{complik}), is inaccessible. The derivative of the logarithm of the likelihood function, $b$ by $b$, ignores the projection data multiplicative factor and reduces to the concise
\begin{equation} \label{partial1}
{{\partial \log f(x,t ; \lambda)} \over {\partial \lambda_b}}= -T + {{x_b} \over {\lambda_b}}
\end{equation}
from which
\begin{equation} \label{partial2}
{{\partial Q(\lambda,\lambda^{(0)})} \over {\partial \lambda_b}} =  -T+{1 \over \lambda_b} E[X_b|n^* ; \lambda^{(0)}]
\end{equation}

This E-step two-parameter function serves the double role of defining the very intuitive M-step iterate $\lambda_b^{(1)} = {1 \over T} E[X_b|n^* ; \lambda^{(0)}]$ as well as identifying (via (\ref{Fisher25})) the incomplete data score function.

As for the conditional expectation,

\begin{eqnarray}
& & E[X_b|n^* ; \lambda^{(0)}]=E[\sum_d t_{b,d}|n^* ; \lambda^{(0)}] \nonumber \\
& = & \sum_{d \neq d_0} E[t_{b,d}|n_d^* ; \lambda^{(0)}] + T \lambda_b^{(0)} q_b = \sum_{d \neq d_0} {{n_d^* \lambda_b^{(0)} p_{b,d}} \over {\sum_{b'} \lambda_{b'}^{(0)} p(b',d)}} +
T \lambda_b^{(0)} q_b \nonumber \\
& = & \lambda_b^{(0)} [T q_b +\sum_{d \neq d_0} {{n_d^* p_{b,d}} \over {g_d^{(0)}}}] \label{condexp}
\end{eqnarray}

The iterate $\lambda_b^{(1)}$ is, as claimed,

\begin{equation} \label{Mstepiterate}
\lambda_b^{(1)} = {1 \over T} E[X_b|n^* ; \lambda^{(0)}] = \lambda_b^{(0)} [q_b + {1 \over T} \sum_{d \neq d_0} {{n_d^* p_{b,d}} \over {g_d^{(0)}}}]
\end{equation}

Completing now (\ref{partial2}),
\begin{equation} \label{partial3}
{{\partial Q(\lambda,\lambda^{(0)})} \over {\partial \lambda_b}} =  -T + {\lambda_b^{(0)} \over \lambda_b} [T q_b + \sum_{d \neq d_0} {{n_d^* p_{b,d}} \over {g_d^{(0)}}}]
\end{equation}

The vector $\lambda^{(0)}$ will be perturbed by the addition of $\epsilon$ only to particular coordinates $\lambda_b^{(0)}$ at a time. Expression (\ref{partial3}) is evaluated at voxel $b_1$ and at each $b_2 \neq b_1$, with all other terms $\lambda_b^{(0)}$ left intact. This will lead to row $b_1$ of the approximate Fisher information matrix.

The term $-T$ can be ignored since only the difference of two versions of (\ref{partial3}) will be used. The ratio ${\lambda_b^{(0)} \over {\lambda_b}}$, that is either ${\lambda_{b_2}^{(0)} \over {\lambda_{b_2}}}=1$ or ${{\lambda_{b_1}^{(0)}+\epsilon} \over {\lambda_{b_1}+\epsilon}}=1$, can be ignored too. With that said, the terms $T q_{b_m}$ can also be ignored. So the relevant part of (\ref{partial3}) is $\sum_{d \neq d_0} p(b_m,d) {{n_d^*} \over {g_d^{(0)}+p(b_1,d) \epsilon}}$, for $m=1$ or $2$. The finite-difference second derivative is then

\begin{equation}
{1 \over {2 \epsilon}}[\sum_{d \neq d_0} p(b_m,d) {{n_d^*} \over {g_d^{(0)}+p(b_1,d) \epsilon}}-\sum_{d \neq d_0} p(b_m,d){{n_d^*} \over {g_d^{(0)}-p(b_1,d) \epsilon}}]=- \sum_{d \neq d_0} n_d^*{{p(b_1,d)p(b_m,d)} \over {(g_d^{(0)})^2 - p(b_1,d)^2 \epsilon^2}}
\end{equation}
as claimed, by letting $\epsilon \rightarrow 0$.

\bigskip

EM-aided differentiation is intended to be a numerical approximation method, where $\epsilon$ needs calibration, perhaps $b$ by $b$. In the current PET application, it led to a closed-form solution. EM-aided differentiation can deliver the observed Fisher information wherever the EM algorithm is applied. This report has illustrated its use for the original 1982 implementation of EM to PET scan by Shepp \& Vardi. Its extension to other variants of the PET model and to other EM-based imaging methods (or otherwise) is left for future research. For more details on numerical implementation,
consult Xu, Baines and Wang (2014).

\bigskip

The following two sections illustrate by examples that inversion of small minors of the Fisher information matrix may provide accurate enough approximations of covariances, but the proper choice of size needs dedicated attention.

\section{A $10\times10$-tissue numerical example} \label{Riklinexample}

\begin{figure}[hbt!]
\begin{center}
{\includegraphics[width=6in,height=3in]{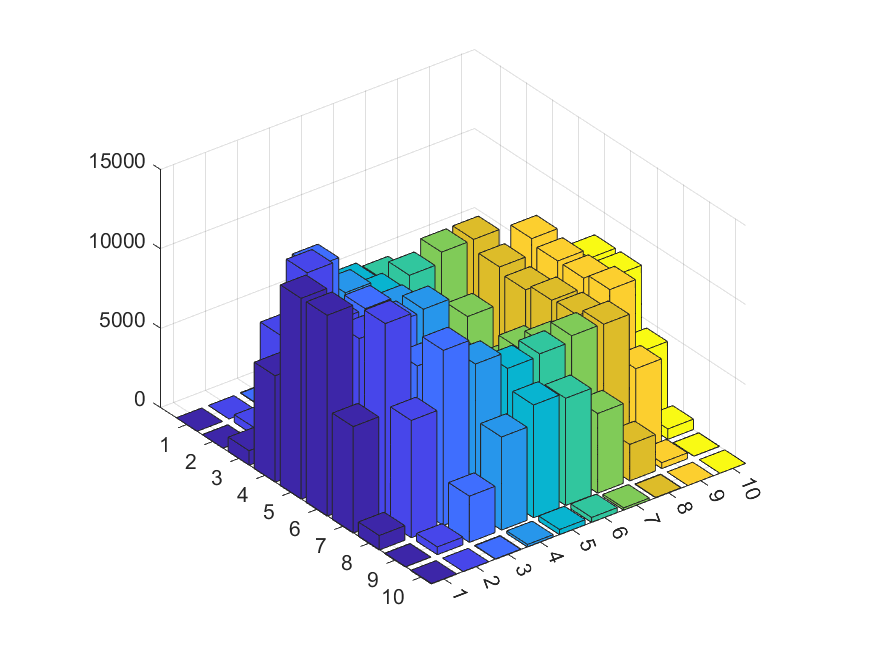}}
\end{center}
\caption{A simulated tumor on a $10 \times 10$ tissue environment}
\label{fig1}
\end{figure}

\begin{figure}[hbt!]
\begin{center}
{\includegraphics[width=2.75in,height=2.5in]{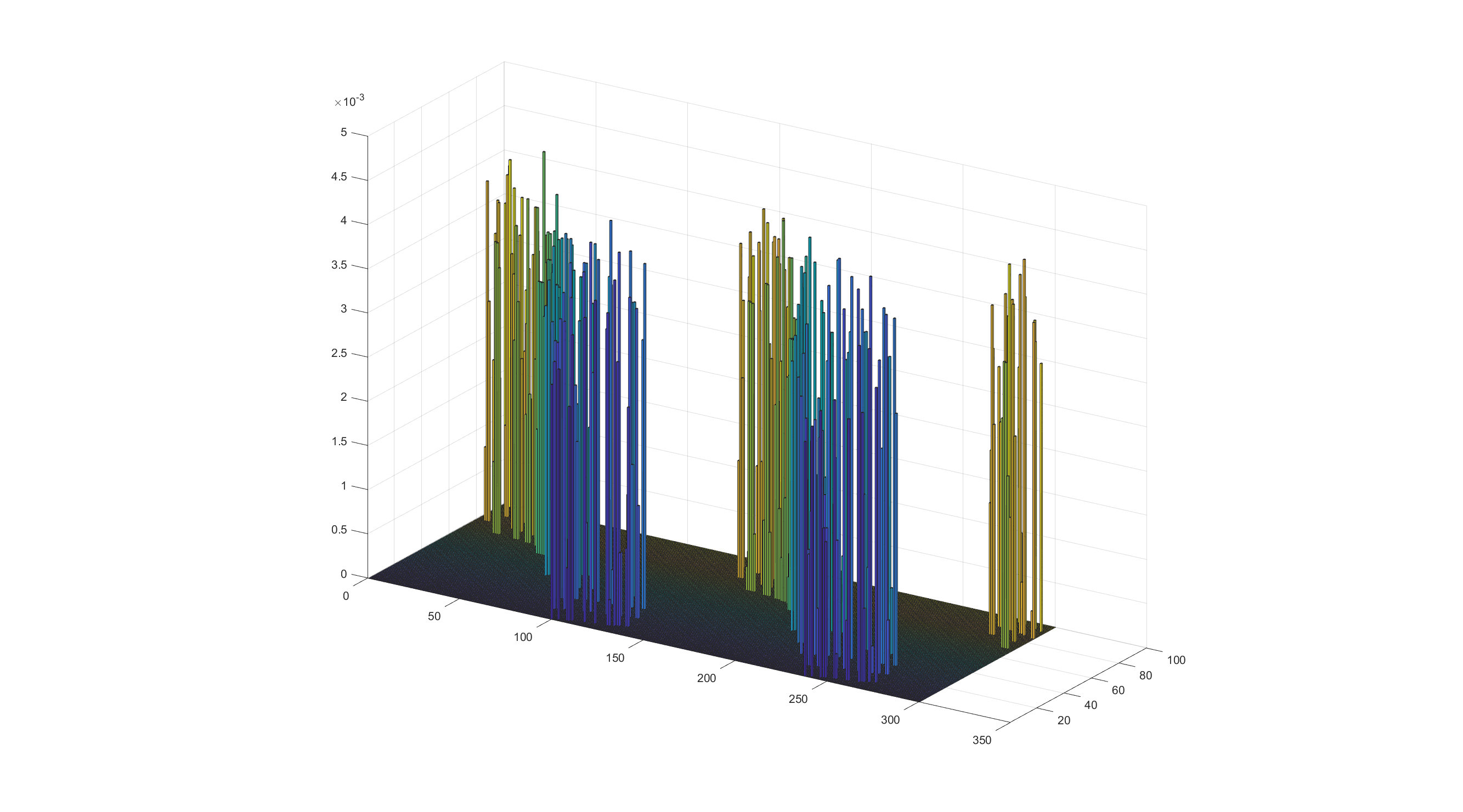}}
\qquad
{\includegraphics[width=2.75in,height=2.5in]{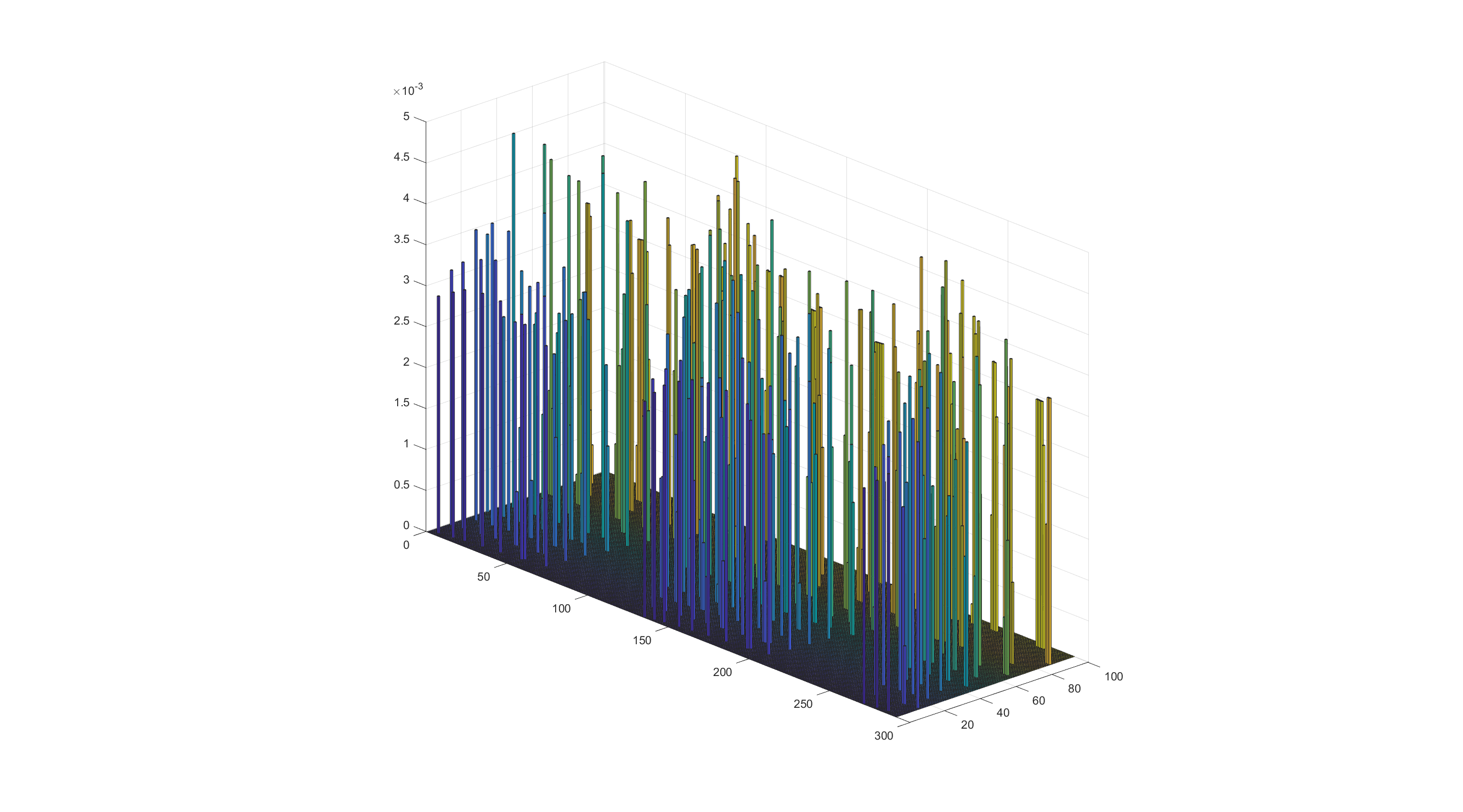}}
\end{center}
\caption{Two $300$-row samples of the $27516$-row system matrix of the example by Riklin, with the 100 voxels linearly ordered. The tumor radiotracer distribution is displayed in Figure \ref{fig1}.}
\label{figsystemmatrix}
\end{figure}

\begin{figure}[hbt!]
\begin{center}
{\includegraphics[width=6in,height=3in]{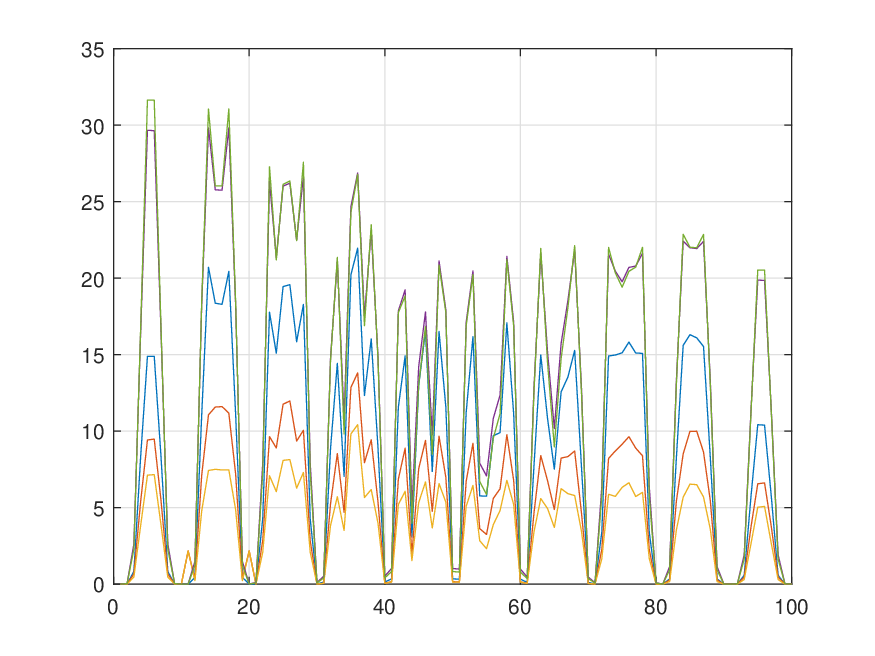}}
\end{center}
\caption{The two almost identical top curves (green and magenta) are the covariance matrix diagonal divided by $500$, and the tumor radiotracer $\lambda$ divided by $400$: in this example, radiotracer and variance of its estimate are proportional to each other. The three bottom curves are (the all-positive and coherently ordered) excesses of the covariance matrix diagonal over its size-$1$ (blue), $5$-tile (red) and $9$-tile (yellow) Fisher minor inversion approximations. These four versions of the diagonal of the covariance matrix come out practically identical, with (displayed) small differences proportional to the variance and radiotracer. The $5$-tile around voxel $i$ appends to $i$ the four geometrically closest neighbors (top, bottom, left, right) and
the $9$-tile around voxel $i$ appends the eight geometrically closest neighbors (the full square). The $9$-tile slightly outperforms the $5$-tile, which reduces variance excess to about half the excess of the $1$-tile.
}
\label{figriklin}
\end{figure}

\bigskip

 This example was provided by Igal Riklin. Figure \ref{fig1} displays the radiotracer distribution, Figure \ref{figsystemmatrix} displays small sections of the system matrix and Figure \ref{figriklin} illustrates the proposed method for approximating the covariance matrix of small hypercubes of voxels, letting the surrounding minor for inversion include some geometrically closest voxels. In this example the radiotracer estimates are quite uncorrelated, and the $5$- and $9$-tile approximants to the variances are adequate. An attempt to enlarge the neighborhoods by appending voxels with high Fisher information with the incumbent voxel, failed to improve performance.
This is a subject for further study, supported by real-life examples, preferably in cooperation with the industry. Likewise, while for this small illustrative problem
a simple Matlab program works well, the potential use of the method for PET scan calibration, the determination of exposure time and hypotheses testing on, e.g., radiotracer differences between tumor sections, needs its incorporation into the established programs.

\begin{figure}[hbt!]
\begin{center}
{\includegraphics[width=6in,height=3in]{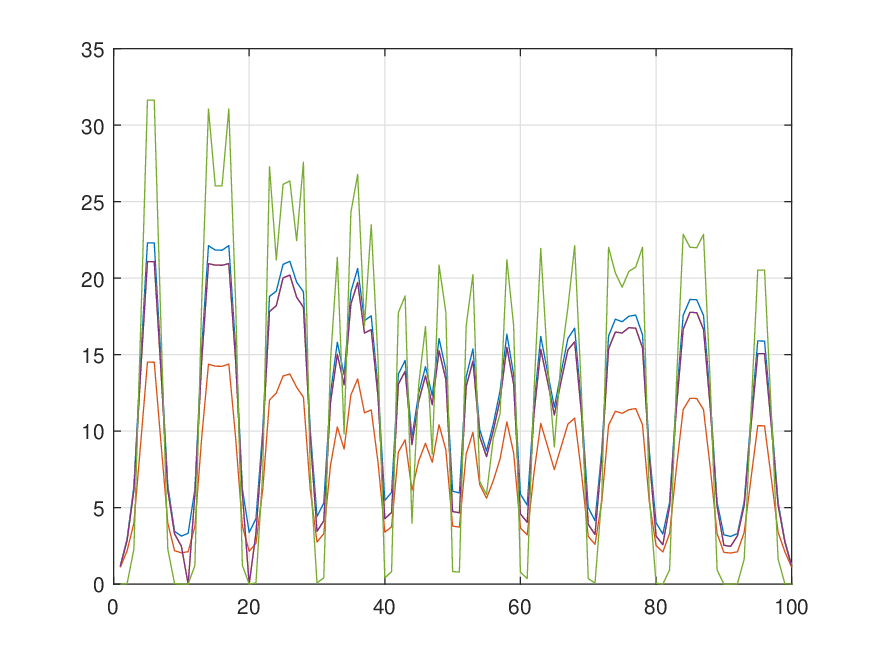}}
\end{center}
\caption{The system matrix modified by averaging. The top curve (green) is the tumor radiotracer $\lambda$ divided by $400$, same as in Figure \ref{figriklin}. The other curves are the covariance matrix diagonal (blue), the $5$- and $9$-tile Fisher minor inversion approximations (magenta) and the size-$1$ version (red). These three versions of the diagonal of the covariance matrix come out proportional but not identical. The $5$-tile provides a good approximation to the covariance matrix, but size-$1$ inversion does not. While the shape of variance and radiotracer graphs is still similar, variances stay away from zero.
}
\label{figriklinsmooth}
\end{figure}

Riklin's system matrix $p$ was smoothed by first replacing each column by the average of that column (weight $0.7$) with the two adjacent columns (weight $0.15$ each), and then averaging each column (weight $0.8$) with the uniform distribution (complementary weight $0.2$). The radiotracer was left untouched. Under this modified system matrix the covariance matrix, the size-$1$ and the $5$- or $9$-tile approximants are different, but the $9$-tile approximant variances are just a trifle above those of the $5$-tile (the two are indistinguishable in the figure). As Figure \ref{figriklinsmooth} shows, while the general shape of variance and radiotracer graphs is still similar, variances stay away from zero. The radiotracer has been divided by $400$ as in Figure \ref{figriklin}, and the variances (now bigger) by $2400$, rather than by $500$. Notice that while Figure \ref{figriklin} displays the small differences between the Fisher diagonal and its approximants, Figure \ref{figriklinsmooth} displays the matrices themselves.

\bigskip

The method was tried on various modifications of the example in Figures \ref{fig1} and \ref{figsystemmatrix} provided by Igal Riklin, with partial analysis illustrated in Figure \ref{figriklin}. Besides the single runs on which the observed Fisher information matrix was calculated, PET was simulated an additional 1000 times, for the purpose of recording the empirical covariance matrix of the MLE. There is very close agreement between the theoretical and empirical versions of the information matrix.

\section{A small numerical example} \label{smallexample}

\begin{figure}[hbt!]
\begin{center}
{\includegraphics[width=2.75in,height=2.5in]{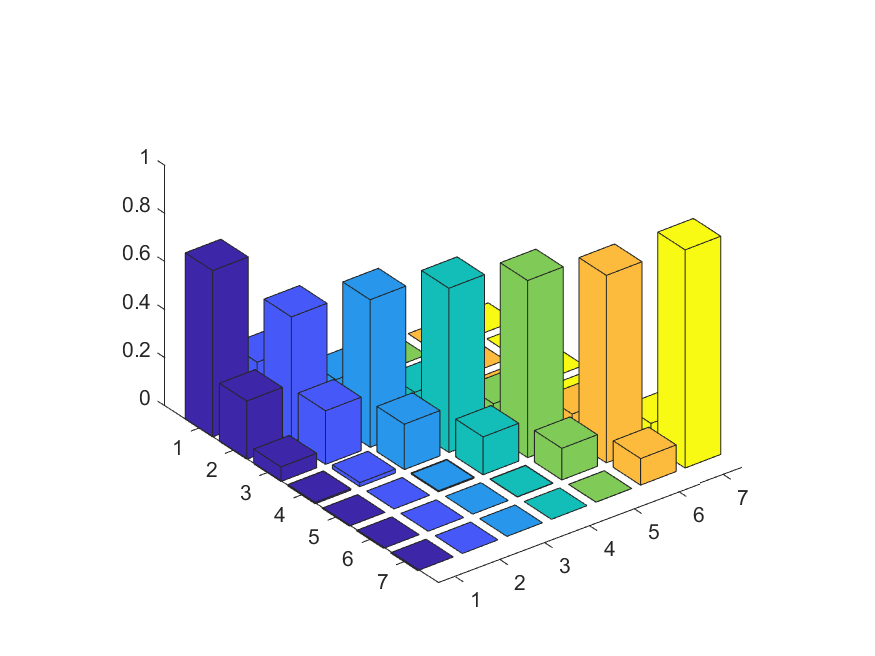}}
\qquad
{\includegraphics[width=2.75in,height=2.5in]{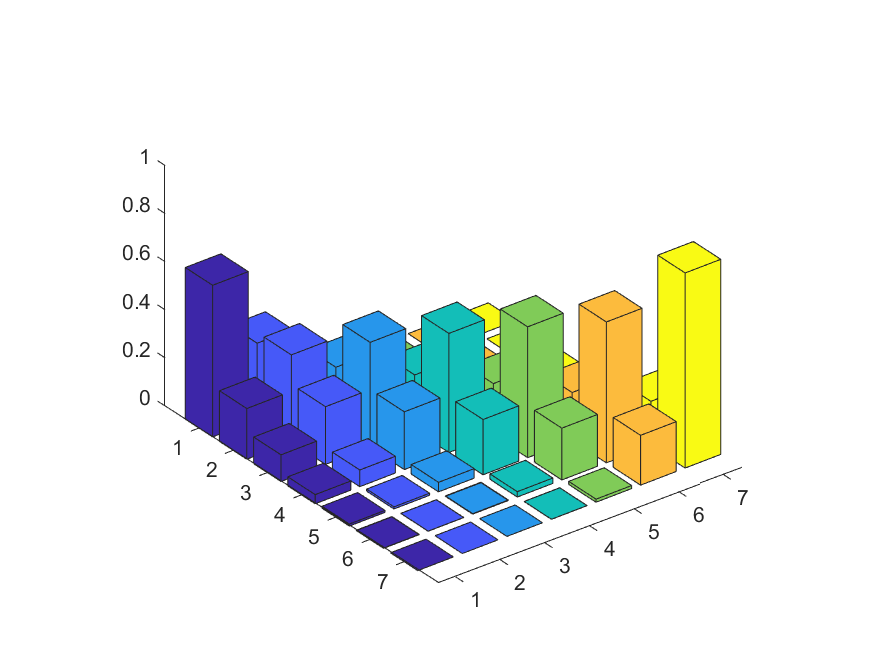}}
\end{center}
\caption{Probability matrices, $\sigma=1$ (left) and $\sigma=1.5$ (right). While similar to the eye, the scenario $\sigma=1.5$ requires $11$ times the sample size to achieve similar accuracy to $\sigma=1$.}
\label{figprob}
\end{figure}

The tissue landscape consists of seven voxels, in which the tumor radiotracer values are $[1 \ 2 \ 3 \ 4 \ 3 \ 2 \ 1]$. There are as many detectors $d$ (playing the role of detector pairs) as voxels $b$, and the probability that an annihilation event at voxel $b$ will be detected at detector $d$ is modelled as the (properly edge-corrected) discretized normal density $\Phi({{d-b+{1 \over 2}} \over {\sigma/\sqrt{b}}})-\Phi({{d-b-{1 \over 2}} \over {\sigma/\sqrt{b}}})$,  where $\Phi$ is the standard normal cumulative distribution function. The term $\sqrt{b}$ introduces some heterogeneity in the splitting quality of voxels, and the term $\sigma$, taken as $1$ and as $1.5$, introduces some global difference in detection quality. Figure \ref{figprob} displays the two probability matrices.

\begin{figure}[hbt!]
\begin{center}
{\includegraphics[width=2.75in,height=2.5in]{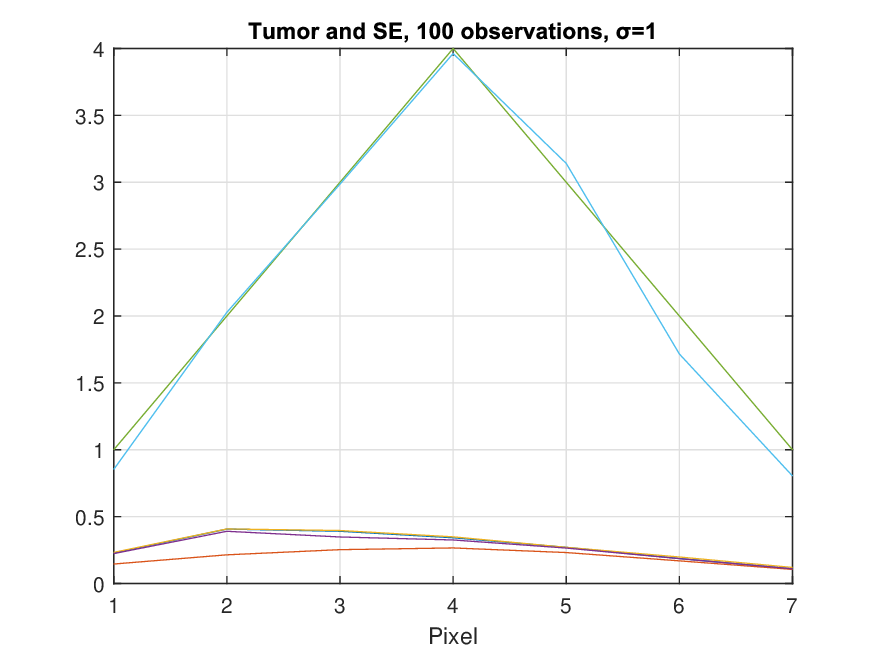}}
\qquad
{\includegraphics[width=2.75in,height=2.5in]{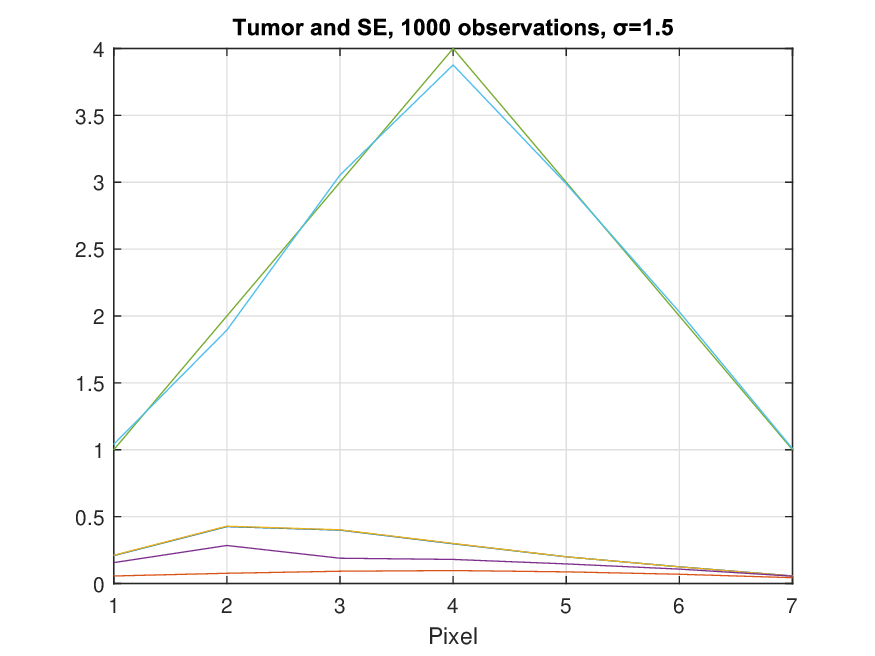}}
\end{center}
\caption{
Radiotracer (green), an instance of its MLE (blue) and its standard error. Empirical (10000 runs) and theoretical SE are practically equal (two indistinguishable curves). $1$-point inversion (red) under-estimates SE, $3$-point inversion (magenta) is almost correct for $\sigma=1$, under-estimate for $\sigma=1.5$}
\label{figtumor}
\end{figure}

Figure \ref{figtumor} displays the tumor, and for each of the two scenarios the tumor radiotracer estimate, based on sample size 100 for $\sigma=1$ and 1000 for $\sigma=1.5$.  The EM algorithm was run for each scenario 10000 times, and the empirical covariance matrices of the MLE were recorded, as well as the observed Fisher information for one of the runs. Figure \ref{figtumor} displays the empirical and observed standard errors, practically indistinguishable. The observed standard error at voxel $b$ is the square root of the $b$-diagonal element of the inverse of the $ 7 \times 7$ observed Fisher information. The $3$-point inversion approximates the observed standard error at voxel $b$ by the square root of the middle diagonal element of the inverse of the $ 3 \times 3$ minor of the observed Fisher information centered at $(b,b)$. The $1$-point inversion simply takes ${1 \over {\sqrt{Fisher(b,b)}}}$. These approximations by inverses of minors of the observed Fisher information will have to be studied carefully in real applications.

\bigskip

Table \ref{tableFisher} presents the correlation matrices obtained from the covariance matrices of the MLE, empirical (below the diagonal) and observed (theoretical, above the diagonal). The upper and lower halves come out practically the same at the resolution of two digits. There is a marked difference in the decay rate of correlations as detectors get away from the pertinent voxel.

\begin{table}[hbt!]
\begin{center}
\begin{small}
\begin{tabular}{rrrrrrr|rrrrrrr}
         \hline
1 &-0.75 & 0.42 & -0.17 & 0.05 & -0.01 & 0.00 & 1 & -0.95 & 0.84 & -0.64 & 0.40 & -0.22 & 0.11\\
-0.76 & 1 & -0.67 & 0.28 & -0.09 & 0.02 & 0.00 & -0.95 & 1 & -0.92 & 0.72 & -0.46 & 0.26 & -0.13\\
0.43 & -0.67 & 1 & -0.53 &0.17 & -0.04 & 0.01 & 0.84 & -0.92 & 1 & -0.86 & 0.58 & -0.33 & 0.16 \\
-0.18 & 0.29 &-0.54 & 1 & -0.42 & 0.11 & 0.02 & -0.64 & 0.72 & -0.87 & 1 & -0.79 & 0.48 & -0.24 \\
0.07 & -0.10 & 0.18 & -0.42 & 1 & -0.34 & 0.07 & 0.40 & -0.45 & 0.58 & -0.79 & 1 & -0.73 & 0.39 \\
-0.03 & 0.02 & -0.04 & 0.09 &-0.33 & 1 & -0.27 & -0.22 & 0.23 & -0.32 & 0.47 & -0.73 & 1 & -0.64 \\
0.01 & 0.01 & 0.00 & -0.02 & 0.07 & -0.27 & 1 & 0.11 & -0.12 & 0.16 & -0.24 & 0.39 & -0.65 & 1 \\
\hline
\end{tabular}
\caption{Empirical (10000 runs) MLE correlation matrix below the diagonal, observed (theoretical) MLE correlation matrix (sample size 100, $\sigma=1$ on the left, sample size 1000, $\sigma=1.5$ on the right) above the diagonal. MLE in voxels at odd distance are negatively correlated, at even distance positively correlated, correlation decays with voxel distance. Empirical and observed correlations are practically equal. The decay is at very different rates in the two scenarios.
}
\label{tableFisher}
\end{small}
\end{center}
\end{table}

\begin{figure}[hbt!]
\begin{center}
{\includegraphics[width=2.75in,height=2.5in]{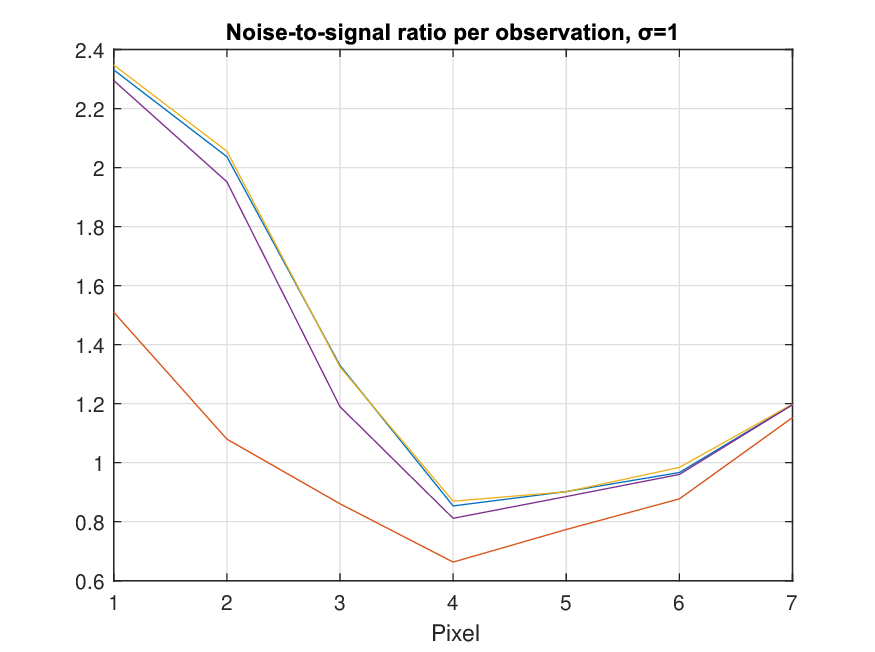}}
\qquad
{\includegraphics[width=2.75in,height=2.5in]{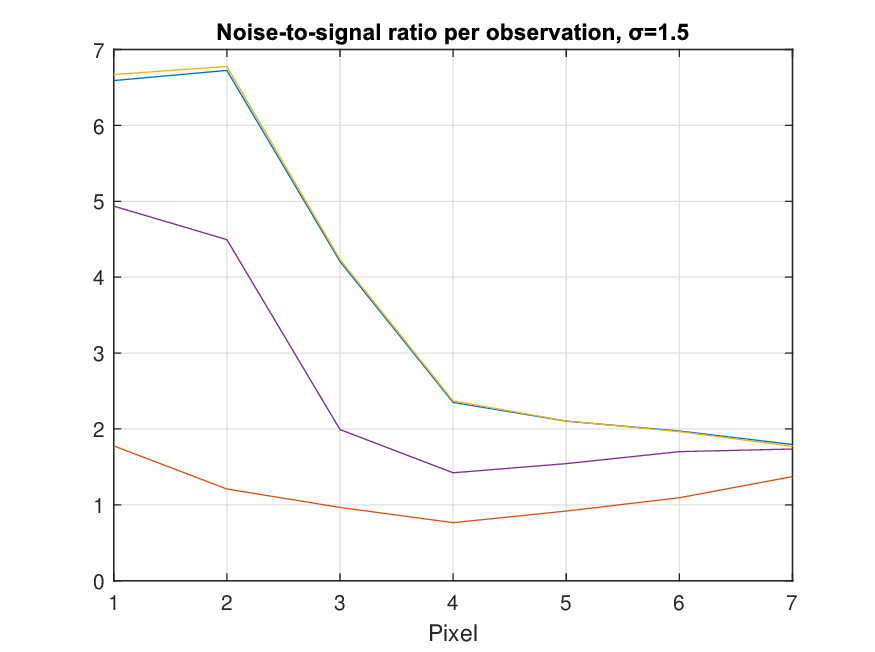}}
\end{center}
\caption{
Noise-to-signal ratio per observation $6.7$ at voxel $2$ for $\sigma=1.5$ (right) is $3.35$ times the N-t-s ratio $2$ for $\sigma=1$ (left). For similar accuracy, sample size should thus be $11$ times higher.
}
\label{fignts}
\end{figure}

\bigskip

Figure \ref{fignts} displays the noise-to-signal ratio, standard error divided by radiotracer value, calibrated so as to be per observation. This ratio is instrumental in determining adequate sample sizes. It shows that the scenario with $\sigma=1$ is much more informative, in the sense that the $\sigma=1.5$-scenario requires 11 times the sample size under $\sigma=1$ to achieve comparable standard errors. The effect of approximating standard errors via small matrix minors is illustrated here as well.

\newpage

\section{Conclusions}

Parametric models in Statistics induce a rigid structure on estimators, via score functions and the Fisher information. Much as non-parametric tools are generally more robust, the rigid structure of parametric inference may carry valuable information waiting to be untapped, as illustrated by Table \ref{tableFisher}. The current study has shown how to obtain in closed form the observed Fisher information matrix of the MLE of the PET scan tumor radiotracer profile. This may indicate how to improve the accuracy of voxel-to-detectors probabilities, help design exposure time and lead to significant discoveries in the tumor topography. It is imperative to speed up convergence. Even for the small $7$-voxel example analyzed, the adequate (but too high) 200 EM iterations for the scenario with $\sigma=1$ are insufficient when detector probabilities are built on $\sigma=1.5$.

\section*{Acknowledgement}

Thanks are due to Igal Riklin for providing a mid-size example on which to try the method. This study was motivated by Riklin's M.Sc. thesis \cite{Riklin}. Thanks are due to Harel Hecht, John Kennedy, Saharon Rosset and David Steinberg for comments and fruitful discussion,
and to Leonid Mendiouk for ongoing attempts to attract the attention of the industry.
The author gratefully acknowledges funding by the ISRAEL SCIENCE FOUNDATION (grant No. 1898/21).

\end{document}